# Merging the Astrophysics and Planetary Science Information Systems

Michael J. Kurtz, Alberto Accomazzi, Edwin A. Henneken

*Harvard-Smithsonian Center for Astrophysics, 60 Garden St, Cambridge MA 02138, USA*
*{mkurtz,aaccomazzi,ehenneken}@cfa.harvard.edu*

## SUMMARY

Conceptually exoplanet research has one foot in the discipline of Astrophysics and the other foot in Planetary Science. Research strategies for exoplanets will require efficient access to data and information from both realms. Astrophysics has a sophisticated, well integrated, distributed information system with archives and data centers which are interlinked with the technical literature via the Astrophysics Data System (ADS). The information system for Planetary Science does not have a central component linking the literature with the observational and theoretical data. Here we propose that the Committee on an Exoplanet Science Strategy recommend that this linkage be built, with the ADS playing the role in Planetary Science which it already plays in Astrophysics. This will require additional resources for the ADS, and the Planetary Data System (PDS), as well as other international collaborators.

## INTRODUCTION

Since the discovery of *51 Peg b* in 1995 work on exoplanets has grown rapidly: currently about 6% of all refereed astronomy articles contain the word "exoplanet" (fig 1). Over this time, research in exoplanets has changed from the purely astronomical (orbits and masses) to include discussion of the properties of the actual objects (planets) discovered. As an example, in 2017 there were 218 papers with the word "atmosphere" in the title and "exoplanet" in the abstract.

Studying planets as objects clearly moves this research from pure Astrophysics toward a more multidisciplinary approach. The primary related fields are Planetary Science (typically represented in journals such as *Icarus*, *Planetary & Space Science*) and geophysics (*JGR Atmospheres*, *J. Atmospheric Sci.*). Eventually biology (e.g. *AstroBiology*) becomes an important research component. About 20% of the recent cited literature in the 2017 papers on exoplanet atmospheres were to journals from these three fields, covering a very broad range of subject matters (fig 2) which are indicative of the ongoing research.

Astrophysics has long had a data and information system which is the envy of other disciplines. Observational data archives (Chandra, HEASARC, IRSA, MAST, ESO), object databases (IPAC/NED, CDS/SIMBAD), curated repositories (CDS/Vizier, CADC), are all connected via the science data they support as represented in the literature. The central vehicle for organizing this information is the ADS (Accomazzi et al. 2016), which allows NED and SIMBAD object searches to be combined with sophisticated text queries and filtering based on data collections.

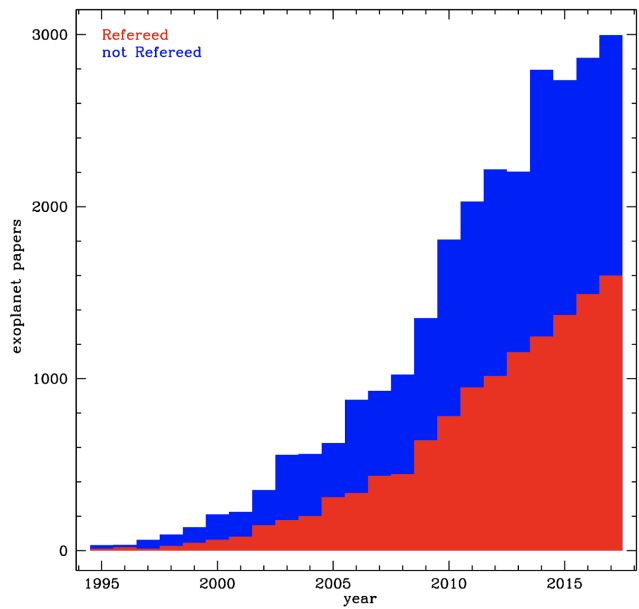

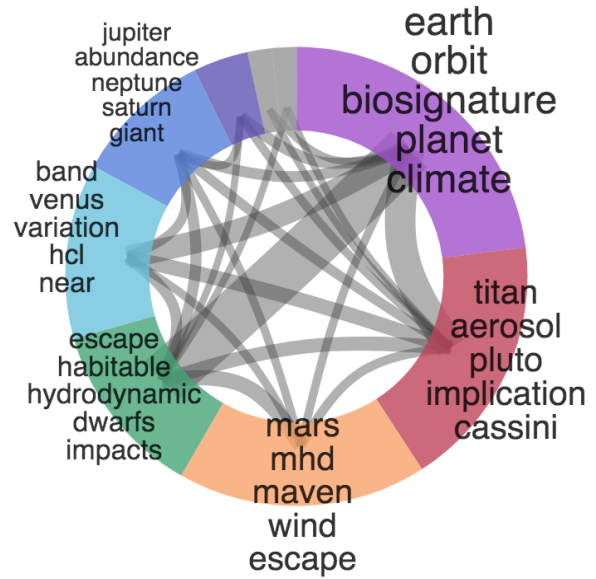

Fig 1. The number of articles mentioning the word "exoplanet" since the discovery of 51Peg b. Currently 6% of all refereed astronomy articles contain the word "exoplanet."

Fig 2. A subject matter clustering of recent cited literature from the 2017 papers discussing atmospheres of exoplanets.

An outcome of having such an ecosystem is that the majority of refereed astronomy articles in ADS have links to data products and data services (70% of *ApJ* articles from 2017), providing straightforward access to multi-wavelength science data analyzed in the papers. This system is very heavily used by professional astronomers, and is routinely expanded. For example, in 2016 the ADS established links to the NASA Exoplanet Science Institute's database (NExScI).

Planetary Science (PS) also has a sophisticated system for data storage and dissemination. The several PDS nodes in the US and the European Planetary Science Archive (PSA) and VESPA project form the core of this effort, along with data archives from several countries. These data often form the basis for the research discussed in scholarly articles, but there is no standard methodology for consistently collecting and disseminating this information, nor an system like the ADS to harvest and merge it with other data sources currently indexed by it.

## PROSPECTS AND CHALLENGES

The lack of a central agency to link Planetary Science data with the PS literature presents a substantial technical impediment to the study and exploration of planets both inside and outside our solar system. While building a system to link PS data with PS publications would in itself be a valuable addition to the PS information system, sharing this system with Astrophysics would provide the connective infrastructure necessary to support the interdisciplinary work required for exoplanet research.

Creating a permanent infrastructure shared by two disciplines is an important strategic step. Wandering stars were part of classical astronomy from antiquity until the beginning of the space age. The discovery of extrasolar planets suggests that a re-alignment of the two now distinct disciplines could be desirable. Once it's decided that this interdisciplinary effort is worth pursuing, a roadmap for an initial implementation of an integrated system is straightforward. We expect that additional improvements and enhancements of the system will become possible once the basic infrastructure outlined below is in place.

## AN INITIAL ROADMAP

Work towards what would be an ongoing collaboration between Astrophysics and Planetary Science would be required from both groups. On the Astrophysics side the ADS would have to expand its core mission to include Planetary Science, so as to cover all the relevant literature (details in appendix). On the PS side the several data archives/repositories would have to develop and maintain a curatorial function which creates authoritative lists of research articles and the relevant data products which were used conduct the research discussed in them.

On the Astrophysics/ADS side the expansion would require an additional curator, programmer, and scientist. On the Planetary Science side each data center would have a person responsible for maintaining the bibliography. In Astrophysics this person is typically a librarian from the main institute (e.g. STScI, ESO), or a data curator associated with the archive (e.g. Chandra). The process of maintaining such a bibliography makes use of supervised text-mining efforts (using the ADS API) to provide a curator with a list of candidate papers and corresponding data products or archives which are used in the papers. The typical curation workflow and required effort are described in Accomazzi et al. (2012).

The process described above provides a way to interlink data products with publications in Astrophysics. An additional curation effort is used to create and maintain a comprehensive list of planetary objects/paper relationships, akin to what NED and SIMBAD do for astronomical objects. As far as exoplanets are concerned, they are already being indexed in SIMBAD, and additional work by the NExScI, at the Paris Observatory, and elsewhere, suggest that this may be a nearly solved problem. The more general problem of indexing the planetary literature and associated data by solar system feature (e.g. Mars's *Ophis Planum*) still remains a challenge worth tackling in the future but is beyond the scope of this whitepaper.

Next to Planetary Science, and closely linked to it, the discipline with the greatest influence on exoplanet research is Geophysics. Just listing the titles of the seven disciplinary sections of the *Journal of Geophysical Research* illustrates this in an obvious way: *Atmospheres, Biogeoscience, Earth Science, Oceans, Planets, Solid Earth, Space Physics*. The ADS currently treats the geophysics literature in a similar manner to Planetary Science and physics (see appendix), and we believe that it would be premature for the time being to expand the effort to link geophysics data with the literature, as this is currently unnecessary for exoplanet research. However, we suggest that this possibility be kept under advisement, and revisited at the time of the mid-decadal review in Astrophysics, and the Planetary Science decadal survey.

## DISCUSSION

Cross-disciplinary endeavours, such as the study of extrasolar planets, require the breaching of institutional silos.  The developments suggested here are modest, but of substantial strategic import, in that they represent a formal interconnection of the two disciplines at the information system level.  This is also likely to have substantial impact: information systems such as ADS or Google have over the past couple of decades been a driver of both research and economic innovation.

A simple example can demonstrate the possibilities.  The *EPOXI* mission data is archived at the PDS Small-Bodies Node (SBN).  There are currently 663 papers in ADS which refer to this mission, 208 of which also contain the word "exoplanet."  None of these papers has a link to the SBN archive, whereas many of them have links to the Astrophysics archives.  The extreme example, Meech et al (2011), has links to imaging and spectral data in the UV and optical at HST, the NIR at ESO and the FIR from Herschel.  By virtue of these links, the relevant datasets from the Astrophysics archives can be downloaded and analyzed in a few clicks, while access to *EPOXI* data requires a search of the PDS nodes.

Likely the most important result of including data in the complex ADS based collaborative multipartite network --- concepts, authors, citations, objects --- is that it facilitates the discovery of relevant data by non-specialists.  This is critical for an intrinsically interdisciplinary field like exoplanets, as no single researcher can be a specialist in every necessary field.  This is as important for PS as it is for Astrophysics, as archival use of PS data already surpasses direct use in some areas: for example, a full 82% of the papers using *EPOXI* data do not have the P.I., Mike A'Hearn, listed as an author.

There are a number of other benefits from this initiative; perhaps some will be serendipitous, but many are already well known in the Astrophysics community: use of the ADS has had a substantial impact on the efficiency of research in Astrophysics (Kurtz et al. 2005).  Sophisticated users routinely move from the ADS to the data centers and archives, and back, substantially enhancing discovery.  The existence of this linked network and the trace of its use provides impact measures useful for funders and project managers (Henneken 2015).

## RECOMMENDATIONS

We suggest that the Committee on an Exoplanet Science Strategy recommend that the Astrophysics and Planetary Science information systems work together to better support exoplanet science.  We suggest that as a first step **the ADS include Planetary Science in its core mission**, alongside Astrophysics, and that **the PDS nodes build and maintain bibliographic indices to their data**, similar to those maintained by the NASA Astrophysics Archives.

We further suggest that this is just the beginning of this endeavour, and that the relevant authorities in both Planetary Science and Astrophysics routinely revisit this issue to see what enhancements should be done to further support the study of extrasolar planets.

## APPENDIX

The basic intellectual structure of the ADS collections is like a three color bullseye (fig 3), with the center being Astrophysics. Here we endeavour to be the authoritative source. We try to be fully complete in collecting and indexing the literature, not just the refereed journals, but also books, conferences, reports, PhD thesis, the so called gray literature. Here we put substantial effort into collaborating with outside groups (CDS, NED, MAST, HEASARC, ESO, etc.); we work with data centers and archives to link papers in our database to the raw and reduced data behind them; we maintain a database of name changes (we know, for example, that Marcia J. Lebofsky and Marcia J. Rieke are the same person). Our aim is to achieve complete coverage with essentially 100% precision and recall for content belonging to this collection. Here we take a leadership role.

Next out, the inner ring, are documents which are likely to be used/cited by authors of documents in the bullseye. Here we have nearly every refereed article in physics, optics, geophysics and Planetary Science. We also have many of the larger conference series from the major publishers (e.g. AIP). We do not attempt to curate this content at the same level of the bullseye. We do not seek the kind of close collaborations which we have in the bullseye core.

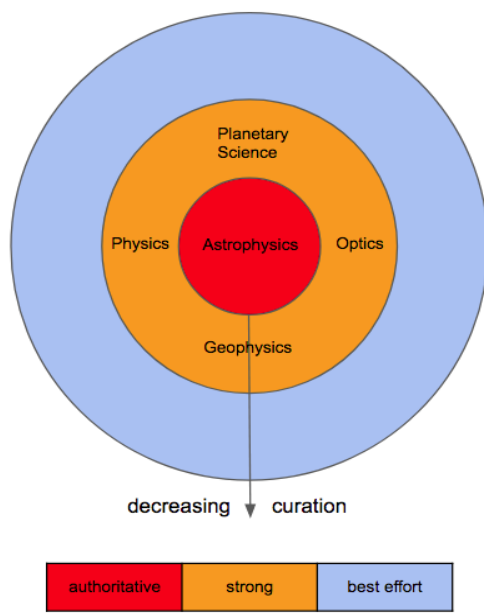

The outer ring are documents which might be used/cited by authors of documents in the inner ring. We only take these if it is very easy, essentially if the publisher provides them to us, or they are available from aggregation systems such as CrossRef. Apart from error checking we perform no curation on these documents.

From the ADS perspective the developments outlined in this white paper would mean moving Planetary Science from the inner ring into the bullseye core. A measure of the effort involved is that there are currently 161,000 citations in the main, refereed PS literature which do not have a corresponding ADS record.

Fig 3. Curation levels of the ADS collections.

From the perspective of the Planetary Science archives these developments would entail expanding or creating their institutional bibliographies to include links to datasets, and sharing these links with the ADS. In all cases in Astrophysics the institutions make substantial internal use of these data (Rots & Winkelman 2015), besides sharing it with ADS. A measure of the scale of this effort is that the Chandra mission maintains data links to nearly 6,000 refereed articles and adds to it one paper per day, on average.